\documentclass[11pt,a4paper]{article}

\usepackage[utf8]{inputenc}
\usepackage[T1]{fontenc}
\usepackage{amsmath,amssymb,amsfonts}
\usepackage{graphicx}
\usepackage{booktabs}
\usepackage{multirow}
\usepackage{hyperref}
\usepackage[margin=1in]{geometry}
\usepackage{authblk}
\usepackage{algorithm}
\usepackage{algpseudocode}
\usepackage{caption}

\title{Automated Classification of Source Code Changes \\ Based on Metrics Clustering \\ in the Software Development Process}

\author[1]{Evgenii G. Knyazev}
\affil[1]{Saint Petersburg State University of Information Technologies, Mechanics and Optics (ITMO University), Saint Petersburg, Russia}

\date{February 2026}

\begin{document}

\maketitle

% ===========================================================================
\begin{center}
\fbox{\parbox{0.92\textwidth}{
\small\textbf{Translator's Note.} This paper is a close English translation of the dissertation abstract (\textit{avtoreferat}) originally written in Russian and defended on December 24, 2009, at Saint Petersburg State University of Information Technologies, Mechanics and Optics (ITMO University), for the degree of Candidate of Technical Sciences (Ph.D.\ equivalent) in specialty 05.13.11 --- Mathematical and Software Support for Computing Systems, Complexes, and Networks.

\medskip
\textbf{Original work:} E.G.~Knyazev, ``Avtomatizirovannaya klassifikatsiya izmeneniy iskhodnogo koda na osnove klasterizatsii metrik v protsesse razrabotki programmnogo obespecheniya'' [Automated Classification of Source Code Changes Based on Metrics Clustering in the Software Development Process], Ph.D.\ dissertation abstract, ITMO University, Saint Petersburg, 2009~\cite{Knyazev2009thesis}.

\medskip
\textbf{Acknowledgment of contributors.} The original research was conducted under the supervision of Prof.\ Anatoly A.\ Shalyto (D.Sc., ITMO University) and in collaboration with Danil G.\ Shopyrin (ITMO University). Key results were co-authored with D.G.~Shopyrin and published in~\cite{Knyazev2008a,Knyazev2008b,Knyazev2007a}. The translation, preparation, and sole responsibility for this arXiv version belong to E.G.~Knyazev.
}}
\end{center}

\bigskip

% ===========================================================================
\begin{abstract}
This paper presents an automated method for classifying source code changes during the software development process based on clustering of change metrics. The method consists of two steps: clustering of metric vectors computed for each code change, followed by expert mapping of the resulting clusters to predefined change classes. The distribution of changes into clusters is performed automatically, while the mapping of clusters to classes is carried out by an expert. Automation of the distribution step substantially reduces the time required for code change review. The k-means algorithm with a cosine similarity measure between metric vectors is used for clustering. Eleven source code metrics are employed, covering lines of code, cyclomatic complexity, file counts, interface changes, and structural changes. The method was validated on five software systems, including two open-source projects (Subversion and NHibernate), and demonstrated classification purity of $P_C = 0.75 \pm 0.05$ and entropy of $E_C = 0.37 \pm 0.06$ at a significance level of 0.05.
\end{abstract}


% ===========================================================================
\section{Introduction}
\label{sec:introduction}

Modern software development organizations work with very large volumes of source code, which complicates its comprehension and analysis and, consequently, hinders quality control. In the process of software quality assurance, source code review (\textit{code review}) plays an important role.

During a code review, the source code is examined in order to detect defects such as algorithmic and architectural errors, violations of adopted coding style, unclear purpose of code fragments, and others. In addition, the reviewer typically searches for unused code, monitors the introduction of redundant or untimely changes, and tracks modifications that could potentially disrupt system functionality or complicate its further development. Source code review allows errors to be detected at early stages of development that would otherwise only be found during the testing phase. In practice, however, source code review usually requires substantial time expenditure.

To simplify code review, reviewers often restrict their attention to the \emph{changes} in the code, since software development usually proceeds iteratively through successive modifications (including the addition of new functionality). Using information about source code modifications simplifies its comprehension by allowing the reviewer to focus attention. Thanks to the widespread use of version control systems, the change history is available for the majority of software projects. However, reviewing changes is typically difficult due to their large number and constraints on the reviewer's time. Therefore, selective review of changes must be conducted, and the criterion for selecting changes may be their membership in a particular class.

It is useful to distinguish classes of changes, such as: implementation of new functionality, removal of unused code, refactoring, bug fixing, and code formatting. Classification of changes is performed not only for understanding the source code, as noted above, but also for assessing the quality of changes based on the code rather than through testing.

Classification is needed to control the development process. For example, if the product has been stabilized, then no changes other than bug fixes should be made. Classification also enables automation of information transfer between participants in the development process. It is used to form the list of changes that require review and, when necessary, to describe them.

This work proposes an automated method for classifying source code changes, consisting of two steps---clustering and mapping of clusters to classes. The distribution of changes into clusters is performed automatically. The mapping of clusters to classes is carried out by an expert. Automation of the distribution of changes into clusters substantially reduces the time required for code change review.

% ===========================================================================
\section{Related Work}
\label{sec:related}

The problem of automating the classification of source code changes has been addressed by many researchers, including A.~Hassan, R.~Holt, S.~Demeyer, S.~Ducasse, S.~Raghavan, R.~Rohana, J.~Maletic, M.~Collard, R.~Robbes, S.~Kim, J.~Whitehead, and others. Methods for classifying changes have been developed based on \textit{heuristic}, \textit{syntactic}, \textit{metric}, and \textit{data mining} approaches to change analysis.

The following methods for automating the classification of source code changes are known:

\begin{enumerate}
    \item \textbf{Heuristic keyword search} in commit comments~\cite{Hassan2004}.
    \item \textbf{Heuristic search and classification of refactorings} based on specific values of defined metrics~\cite{Demeyer2000}.
    \item \textbf{Comparison of syntax trees} of code versions~\cite{Raghavan2004}.
    \item \textbf{Analysis of syntactic differences} between code versions using tags embedded in the code~\cite{Maletic2004}.
    \item \textbf{Version control system storing abstract syntax trees}, obtained from the development environment database~\cite{Robbes2007}.
    \item \textbf{Classification of changes by the presence of possible errors}~\cite{Kim2006}.
\end{enumerate}

The drawback of the first two methods is their incorrect behavior for certain input data, since it is difficult to account for all possible data values when developing heuristics. The drawbacks of the next two methods are their algorithmic complexity, dependence on the programming language, and lack of adaptability. The use of syntactic methods for change classification is justified only for analyzing simple changes. The drawback of the fifth method is the need for redundant storage and processing of the full set of modifications in the version control system performed by the user in the development environment, since not all of them are preserved in the final code. The drawback of the sixth method, based on a data mining approach, is the need to construct a training set containing changes that introduce errors and their corrections. Constructing such a set is a non-trivial task. Moreover, the majority of the listed methods share a common drawback---they are designed to build a single, fixed classification and do not allow reconfiguration for other classifications.

These drawbacks can be overcome by combining the metric approach with a clustering method that does not require the formation of a training set. Therefore, developing a method for automated classification of changes using clustering and the metric approach is a relevant research task.

% ===========================================================================
\section{Research Objectives}
\label{sec:objectives}

The \textbf{goal} of this dissertation is to develop a method for automated classification of source code changes during the software development process based on metrics clustering.

The \textbf{main research tasks} are:

\begin{enumerate}
    \item To justify the feasibility of partial automation of the classification of source code changes through metrics clustering.
    \item To develop a method for automated classification of source code changes based on clustering of change metrics.
    \item To deploy the results in the practice of software development.
\end{enumerate}

\subsection{Scientific Novelty}

The following results, possessing scientific novelty, are presented for defense:

\begin{enumerate}
    \item Justification of the feasibility of partial automation of source code change classification through the metrics clustering method, by formulating a hypothesis about automated classification and its experimental confirmation.
    \item Justification of the choice of the \textit{k-means method with a cosine-based proximity measure} for clustering change metric vectors.
    \item A method for automated classification of source code changes based on clustering of change metrics, enabling a reduction in the number of changes that must be classified manually.
\end{enumerate}

% ===========================================================================
\section{Investigating the Feasibility of Automated Classification}
\label{sec:feasibility}

\subsection{Formal Framework}

Consider a software system $P$ evolving over time. The state of this system at each moment $t$ is given by its current code $S_t$. For convenience, let us denote the sets of unchanged states $S_t$ during consecutive time intervals $t \in (t_{r-1}, \ldots, t_r)$ as $S_r$, where $r$ is an integer, $1 \le r \le N$, and $N$ is the total number of distinct code states. A \textit{source code change} is defined as a mapping $\delta_r$ that transforms the code from the preceding state $S_{r-1}$ to the modified state $S_r$:
\begin{equation}
    S_{r-1} \xrightarrow{\delta_r} S_r.
\end{equation}

Each change $\delta$ may be assigned to a class $c$, where $c \in C$, and $C$ represents the set of change classes specific to the software system being developed. The most common change classes are: implementation of new functionality, bug fixing, refactoring, removal of unused code, and code formatting.

The task of assigning a change to a class is labor-intensive and requires the participation of a highly qualified expert. The expert determines the composition of the set of classes, specific to each software system, and also decides which class a particular change belongs to.

The result of expert classification depends not only on the software system being analyzed and the set of changes, but also on the particular expert. Therefore, when different experts classify the same changes, different class distributions are possible.

We define the \textit{verification set} $\Delta_e$ as a set of changes classified by an expert. To eliminate possible contradictions, in this work verification sets are constructed by two experts, and changes classified differently by them are excluded from the verification set.

\subsection{Hypothesis}

The method of clustering change metrics makes it possible to mitigate the subjectivity of classification and, as demonstrated in this dissertation, can provide partial automation of source code change classification. The automation of classification is based on the following hypothesis.

\begin{quote}
\textbf{Hypothesis 1.} \textit{Automated classification of code changes for a given software system is feasible using metrics clustering.}

\textit{Let a set of change classes be given and it be required to classify the set of changes of a software system. Let also a metric vector be constructed for each change. Then these metric vectors can be clustered such that each resulting cluster corresponds to one class of changes.}
\end{quote}

% ===========================================================================
\section{Method for Automated Classification of Source Code Changes}
\label{sec:method}

The proposed method builds the automation of source code change classification on the basis of clustering change metric vectors. The method schema is presented in Figure~\ref{fig:method}. The input data consist of: the set of changes to classify $\Delta = \{\delta_r\}$, the set of expert classes $C = \{c_i\}$ ($1 \le i \le n$), the set of change metrics $\boldsymbol{\mu}$, and the number of clusters $k$, which is initially set equal to the number of expert classes $n$. The output data are the set of classified changes.

\begin{figure}[ht]
\centering
\fbox{\parbox{0.85\textwidth}{
\begin{enumerate}
    \item \textbf{Expert configuration}: select the set of change metrics $\boldsymbol{\mu}$ to ensure given levels $P_{C_{min}}$, $E_{C_{max}}$ of classification quality criteria. Set $k = n$.
    \item \textbf{Compute metric vectors} for all changes.
    \item \textbf{Cluster} the metric vectors.
    \item \textbf{Check} clustering quality functional $I > I_{min}$.
    \begin{itemize}
        \item If not satisfied: increment $k \leftarrow k + 1$ and return to step 3.
    \end{itemize}
    \item \textbf{Expert mapping}: map each cluster to an expert class.
    \item \textbf{Automatic classification}: classify all changes based on the cluster-to-class mapping.
    \item \textbf{Evaluate} quality criteria $P_C$, $E_C$ of the method.
\end{enumerate}
}}
\caption{Schema of the method for classifying source code changes based on metrics clustering.}
\label{fig:method}
\end{figure}

\subsection{Computing Change Metric Vectors}
\label{sec:metrics}

For each change, a vector is formed from the values of the metrics selected during the method configuration stage. We define a \textit{metric of source code change} $\delta_r$ as a numerical quantity $\mu\delta_r$ that characterizes the given change. This metric for code $S$ is computed as:
\begin{equation}
    \mu\delta_r = M^{\delta}(S_{r-1},\, S_r),
\end{equation}
where $M^{\delta}$ is the change metric. In practice, it is more convenient to compute metrics in terms of added, modified, and deleted lines of code.

We generalize the notion of a code line $l$ as a sequence of lexemes $s_1 s_2 \ldots s_e$, where $s_e$ is the end-of-line lexeme. Each change $\delta_r$ can be represented as a collection of added lines $L_+$, deleted lines $L_-$, and modified lines $L^*$:
\begin{equation}
    \delta_r = \{L_+,\, L_-,\, L^*\},
\end{equation}
where $L^*$ is the set of pairs $\langle l^*_-, l^*_+ \rangle$, with $l^*_-$ being the line before the change and $l^*_+$ the modified line. Such a representation of changes is produced by the \textit{edit script} method. An edit script is a sequence of insert, delete, and replace operations needed to transform one string into another in the simplest way.

As an example, we present formulas for computing metrics based on \textbf{cyclomatic complexity}. Cyclomatic complexity is a metric designed to evaluate the complexity of a program's control flow. Define the function $CC(l)$ computing the simplified cyclomatic complexity of a code line $l$ as the number of language constructs controlling the program execution flow found in line $l$:
\begin{equation}
    CC(l) = \sum_{s_i \in l} [s_i \in S_{CF}], \quad 1 \le i \le n_l,
\end{equation}
where $\{s_i\}$ is the set of lexemes in line $l$, $n_l$ is the number of lexemes in line $l$, $S_{CF}$ is the set of lexemes representing control flow constructs. The expression $[B]$ takes the value one when predicate $B$ is true and zero when $B$ is false.

The formulas for computing the cyclomatic complexity metrics of added code $CC_+$, deleted code $CC_-$, and modified code $CC^*$ are:
\begin{align}
    CC_+ &= \sum_{l_+ \in L_+} CC(l_+), \\
    CC_- &= \sum_{l_- \in L_-} CC(l_-), \\
    CC^* &= \sum_{\langle l^*_-, l^*_+ \rangle \in L^*} \bigl(CC(l^*_+) - CC(l^*_-)\bigr).
\end{align}

Table~\ref{tab:metrics} lists all eleven metrics used in this dissertation.

\begin{table}[ht]
\centering
\caption{Source code change metrics used in the dissertation.}
\label{tab:metrics}
\begin{tabular}{@{}clcl@{}}
\toprule
\# & Metric Description & \# & Metric Description \\
\midrule
1. $LOC_+$ & Lines of code added       & 7. $F^*$ & Files of source code modified \\
2. $LOC_-$ & Lines of code deleted     & 8. $I_+$ & Interfaces added \\
3. $LOC^*$ & Lines of code modified    & 9. $I_-$ & Interfaces deleted \\
4. $CC_+$  & Cyclomatic complexity of added code   & 10. $CS_+$ & Classes and structures added \\
5. $CC_-$  & Cyclomatic complexity of deleted code  & 11. $CS_-$ & Classes and structures deleted \\
6. $CC^*$  & Cyclomatic complexity of modified code & & \\
\bottomrule
\end{tabular}
\end{table}

For each change $\delta_r$, a $p$-dimensional metric vector is computed, where $p$ is the number of metrics used:
\begin{equation}
    \boldsymbol{\mu}\delta_r = \langle \mu_1\delta_r,\; \mu_2\delta_r,\; \ldots,\; \mu_p\delta_r \rangle.
\end{equation}

\subsection{Clustering of Change Metric Vectors}
\label{sec:clustering}

At this stage, a set of change clusters $Q$ is formed. Clusters $q_j \in Q$ are non-overlapping subsets of changes grouped by a given proximity criterion during the clustering process.

Clustering is performed using the \textbf{k-means algorithm} with a \textbf{cosine proximity measure} between changes. The choice of the k-means clustering algorithm is motivated by its simplicity of use and acceptable quality of results. The use of a cosine-based proximity measure addresses the problem of k-means sensitivity to the scale of individual variables:
\begin{equation}
    \rho(v_1, v_2) = \cos(v_1, v_2) = \frac{v_1^{\top} v_2}{\|v_1\| \cdot \|v_2\|},
    \label{eq:cosine}
\end{equation}
where $v_1$, $v_2$ are change metric vectors and $v_1^{\top}$ is the transpose of $v_1$. Objects are considered more similar when the value of $\rho$ is closer to one.

When Euclidean distance is used, changes are grouped into clusters based on their scale rather than their expert class. By using the cosine-based measure $\rho$, which does not depend on the length of change metric vectors, it becomes possible to distinguish change classes without regard to their scale.

The clustering algorithm proceeds as follows:

\begin{enumerate}
    \item \textbf{Initial partition}: $Q^{(0)} = \{q_1^{(0)}, q_2^{(0)}, \ldots, q_k^{(0)}\}$ of the set of objects $\{\delta_i\}$ such that the $k$ most distant changes are placed first into different clusters:
    \begin{align}
        q_1^{(0)} &= \{\delta_1\}, \\
        q_j^{(0)} &= \{\delta_i \mid \rho(\boldsymbol{\mu}\delta_i, cq_t^{(0)}) = \min_{i,t} \rho(\boldsymbol{\mu}\delta_i, cq_t^{(0)})\}, \quad 2 \le j \le k.
    \end{align}
    \item Set iteration counter $l = 1$.
    \item \textbf{Compute cluster centers}:
    \begin{equation}
        cq_j^{(l)} = \frac{\sum_i [\delta_i \in q_j^{(l-1)}] \cdot \boldsymbol{\mu}\delta_i}{\sum_i [\delta_i \in q_j^{(l-1)}]}, \quad 1 \le j \le k.
    \end{equation}
    \item \textbf{Update cluster assignments}:
    \begin{equation}
        Q^{(l)} = \{q_j^{(l)}\}, \quad q_j^{(l)} = \{\delta_i \mid \rho(\boldsymbol{\mu}\delta_i, cq_j^{(l)}) = \max_j \rho(\boldsymbol{\mu}\delta_i, cq_j^{(l)})\}.
    \end{equation}
    \item \textbf{Check convergence}: $\sum_i \|q_j^l - q_j^{l-1}\| = 0$, where ``$-$'' denotes the symmetric difference $A - B = (A \cup B) \setminus (A \cap B)$. If the condition is satisfied, the process terminates; otherwise, set $l \leftarrow l + 1$ and return to step 3.
\end{enumerate}

This algorithm was implemented using the open-source tool \textit{CLUTO} (Karypis Lab, University of Minnesota, USA).

\subsection{Clustering Quality Criterion}
\label{sec:quality_clustering}

To evaluate the quality of the partition of changes into clusters, the functional $I$ was chosen as the simplest measure of the ``density'' of change groupings within clusters:
\begin{equation}
    I = \sum_{j=1}^{k} \sqrt{\sum_{\delta_1, \delta_2 \in q_j} \rho(\delta_1, \delta_2)},
\end{equation}
where $k$ is the number of clusters, $q_j$ is the cluster with index $j$, $\rho$ is the proximity measure, and $\delta_1, \delta_2$ are changes belonging to cluster $q_j$. The meaning of this functional is that the higher its value, the more densely changes are grouped within clusters, which indicates a higher quality of the clustering solution.

At this stage, the condition $I > I_{min}$ is checked against a minimum level set by the expert. If the condition is not met, the number of clusters $k$ is adjusted.

\subsection{Expert Mapping of Clusters to Classes}
\label{sec:mapping}

At this stage, each cluster is mapped to some expert class. A correspondence is established between clusters $q_1, q_2, \ldots, q_k$ and change classes $c_1, c_2, \ldots, c_n$. The mapping is conducted on the basis of selective expert classification of changes from each cluster.

In each cluster, several changes closest to the center are selected and classified by the expert. Based on the predominance of changes of a given class $c$ in cluster $q$, all objects in cluster $q$ are classified as $c$: $q \rightarrow c$.

\subsection{Quality Criteria of the Method}
\label{sec:quality_method}

The quality of the automated classification method is evaluated on the verification set of changes $\Delta_e$.

For each cluster $q_j$, the following criteria are computed: \textit{purity} $P(q_j)$ and \textit{entropy} $E(q_j)$. Purity $P(q_j)$ of cluster $q_j$ is the ratio of the number of its changes belonging to the expert class $c_i$ corresponding to the given cluster ($q_j \rightarrow c_i$) to the total number of changes from $\Delta_e$ in cluster $q_j$. The averaged values of purity $P_Q$ and entropy $E_Q$ across all clusters are used as quality estimates.

It is preferable to assess the correspondence between automated and expert classification using the alternative criteria of purity $P_C$ and entropy $E_C$ of \textit{the distribution of changes across expert classes}:

\begin{align}
    P_Q(\Delta_e) &= \sum_{j=1}^{k} \frac{n_j^e}{N_e} P(q_j), \quad P(q_j) = \frac{n_j^a}{n_j^e}, \label{eq:purity_q} \\
    E_Q(\Delta_e) &= \sum_{j=1}^{k} \frac{n_j^e}{N_e} E(q_j), \quad E(q_j) = -\frac{1}{\log n} \sum_{i=1}^{n} \frac{n_j^i}{n_j^e} \log \frac{n_j^i}{n_j^e}, \label{eq:entropy_q}
\end{align}
where $N_e$ is the total number of changes in $\Delta_e$, $n$ is the number of change classes, $k$ is the number of clusters, $n_j^e$ is the number of changes from $\Delta_e$ that fell into cluster $q_j$, $n_j^i$ is the number of changes from $\Delta_e$ in cluster $q_j$ belonging to class $c_i$, and $n_j^a$ is the number of changes in cluster $q_j$ from $\Delta_e$ belonging to the expert class $c_j^a$ to which cluster $q_j$ was mapped.

We use criteria $P_Q$ and $E_Q$ to reformulate Hypothesis~1 as follows:

\begin{quote}
\textit{For a verification set of changes $\Delta_e$, the automated method of change classification produces a partition of changes into clusters with subsequent mapping to their classes such that, for given levels $P_{Q_{min}}$, $E_{Q_{max}}$, the following inequality holds:}
\begin{equation}
    P_Q(\Delta_E) > P_{Q_{min}}, \quad E_Q(\Delta_E) < E_{Q_{max}}.
\end{equation}
\end{quote}

To verify this hypothesis and to estimate confidence intervals for purity and entropy values, the well-known \textit{bootstrap resampling} method is used, since the experimental data are insufficient for standard statistical estimation. The classification of changes is a process requiring substantial expert time expenditure. Therefore, in practice it is very difficult to construct verification sets of sufficient size.

The set of changes is divided into $M$ equal parts. Then the first part is excluded and the remaining $M-1$ parts are used as a separate sample. The part excluded in the previous step is returned, the second part is excluded, and the remaining parts numbered $1, 3, \ldots, M$ are used as the next sample, and so on. Based on the samples generated in this manner, statistical parameter estimation is performed.

% ===========================================================================
\section{Experimental Validation}
\label{sec:experiments}

\subsection{Application Use Cases}

The first section of Chapter~3 contains a detailed description of use cases for the automated classification of changes in the software development process. Automated change classification can be useful to all members of the development team.

\subsection{Industrial Deployment}

The method was deployed at \textit{Transas Technologies} (Saint Petersburg) for automated classification of source code changes based on metrics clustering. Deployment was carried out (as evidenced by implementation certificates) during the development of: (1) the mobile object monitoring system \textit{Navi-Manager}; (2) a component of the global fleet monitoring system \textit{LRIT} (Long-Range Identification and Tracking); (3) the student action monitoring system \textit{e-Tutor 5000} for ship, crane, and other training simulators.

The proposed method was also used to analyze open-source software systems: the version control system \textit{Subversion} (subversion.tigris.org, CollabNET, USA) and the object-relational mapping framework \textit{NHibernate} (JBoss, USA).

\subsection{Example: NHibernate}
\label{sec:nhibernate}

As an example of using the proposed method, we describe its application during the development of the NHibernate system. The input data for the experiment are presented in Table~\ref{tab:nhibernate_input}.

\begin{table}[ht]
\centering
\caption{Input data for the experiment on classifying changes in the NHibernate system.}
\label{tab:nhibernate_input}
\begin{tabular}{@{}p{7cm}p{7cm}@{}}
\toprule
\textbf{General Information} & \\
\midrule
Expert classes $C$ & Code deletion, new functionality implementation, bug fixing, refactoring, code formatting with comment changes \\
Clustering method for change metrics & k-means \\
Proximity measure for metric vectors $\rho$ & Cosine of the angle between metric vectors \\
Set of metrics for clustering $\boldsymbol{\mu}$ & As listed in Table~\ref{tab:metrics} \\
\midrule
\textbf{Input Data} & \\
\midrule
Total number of changes to classify & 2069 \\
Number of expert-classified changes & 73 \\
\midrule
\textbf{Method Configuration} & \\
\midrule
Selected number of clusters $k$ & 12 \\
\bottomrule
\end{tabular}
\end{table}

The automated classification proceeded as follows:

\begin{enumerate}
    \item During \textbf{expert configuration}, the metrics listed in Table~\ref{tab:metrics} were selected. The verification set $\Delta_{e68}$ of 68 changes was constructed as follows: from each class identified by the automated method, 16 changes were randomly selected, yielding 80 total; of these, 68 changes were retained for which the expert classifications of two independent experts agreed.

    \item \textbf{Metric vectors} were computed for all 2069 changes.

    \item The metric vectors were \textbf{partitioned into twelve clusters}. The choice of $k$ was based on analyzing the clustering quality functional $I$ and the size of the smallest ``bad'' cluster with substantially different change proximity measures.

    \item The \textbf{clustering quality criterion} was evaluated for different numbers of clusters $k$ from 5 to 15. The clustering quality functional $I$ increases with growing $k$. At $k = 12$, the smallest ``bad'' cluster reaches 547 changes.

    \item \textbf{Cluster-to-class mapping} was performed by selecting changes from each cluster, classifying them by expert review, and establishing the correspondence of each cluster to a class based on the predominance of changes of that class.

    \item \textbf{Automatic classification} was performed based on the cluster-to-class mapping from the previous step.

    \item The \textbf{quality criteria} of the method---purity $P_Q(\Delta_{e68})$ and entropy $E_Q(\Delta_{e68})$---for the partition of the verification set $\Delta_{e68}$ across clusters are presented in Table~\ref{tab:clusters}.
\end{enumerate}

\begin{table}[ht]
\centering
\caption{Distribution of verification set changes across clusters. Expert class labels: ``B''---bug fixing, ``F''---code formatting with comment changes, ``D''---code deletion, ``N''---new functionality, ``R''---refactoring. $n_j^e$---number of changes from $\Delta_{e68}$ in cluster $q_j$.}
\label{tab:clusters}
\small
\begin{tabular}{@{}cccccccccc@{}}
\toprule
Cluster $j$ & Mapping $q_j \rightarrow c_i$ & B & F & N & D & R & $n_j^e$ & $P_Q$ & $E_Q$ \\
\midrule
0  & B   & 9 & 0 & 0 & 1 & 1 & 11 & 0.82 & 0.37 \\
1  & F   & 1 & 4 & 0 & 0 & 0 &  5 & 0.80 & 0.31 \\
2  & F   & 0 & 1 & 0 & 0 & 0 &  1 & 1    & 0    \\
3  & F   & 0 & 0 & 0 & 0 & 0 &  0 & 1    & 0    \\
4  & F   & 1 & 7 & 0 & 0 & 0 &  8 & 0.88 & 0.23 \\
5  & N   & 0 & 0 & 8 & 0 & 0 &  8 & 1    & 0    \\
6  & N   & 0 & 0 & 4 & 0 & 0 &  4 & 1    & 0    \\
7  & N   & 1 & 1 & 1 & 0 & 0 &  3 & 0.33 & 0.68 \\
8  & D   & 1 & 0 & 0 & 13& 1 & 15 & 0.87 & 0.30 \\
9  & R   & 2 & 1 & 0 & 0 & 3 &  6 & 0.50 & 0.63 \\
10 & R   & 0 & 0 & 0 & 0 & 1 &  1 & 1    & 0    \\
11 & R   & 2 & 2 & 0 & 0 & 2 &  6 & 0.33 & 0.68 \\
\midrule
\textbf{Total} & & 17 & 16 & 13 & 14 & 8 & 68 & 0.78 & 0.32 \\
\bottomrule
\end{tabular}
\end{table}

Table~\ref{tab:quality} presents the averaged quality criteria $P_C$ and $E_C$. The values were obtained by: first, combining rows for clusters representing the same expert class; second, computing $P_C(\Delta_{e68})$ and $E_C(\Delta_{e68})$ for these combined rows; and third, averaging the values obtained across five experiments using the bootstrap resampling method. The significance level for constructing confidence intervals for $P_C$ and $E_C$ was set at 0.05.

\begin{table}[ht]
\centering
\caption{Averaged quality criteria of automated classification for NHibernate system changes.}
\label{tab:quality}
\begin{tabular}{@{}lr@{}}
\toprule
Number of changes in the resampled verification set & 272 \\
\midrule
Classification quality & $P_C = 0.75 \pm 0.05$ \\
                        & $E_C = 0.37 \pm 0.06$ \\
\bottomrule
\end{tabular}
\end{table}

From Table~\ref{tab:quality}, it follows that for the analyzed software system, on average 75\% of changes will be correctly classified by the proposed method. Furthermore, the obtained entropy value indicates that there is substantial uncertainty in the correspondence of clusters to classes. During the expert cluster-to-class mapping stage, it was found that this is primarily due to insufficient separation of clusters corresponding to the \textit{refactoring} class.

The application of the automated classification method enabled the distribution of 2069 changes into five classes on the basis of clustering into twelve clusters. For comparison with the ``purely expert assessment'' approach, we note the following figures: only 73 changes were classified to map clusters to classes during the expert classification process. The automated classifier correctly assigned 53 changes out of the 68-element verification set to the same classes as the experts. Consequently, the application of the method made it possible to \textbf{substantially reduce the time experts spent} on classifying changes.

For the analyzed software system, \textbf{Hypothesis~1 is confirmed} for the following values of $P_{Q_{min}}$ and $E_{Q_{max}}$ of automated classification at a significance level of 0.05:
\begin{equation}
    P_{Q_{min}} = 0.71, \quad E_{Q_{max}} = 0.34,
\end{equation}
which indicates acceptable quality of automated classification for practical application of the method.

Based on the experiments conducted, it can be concluded that the method works well when uniform changes are made. When changes combine heterogeneous code modifications, the quality of automated classification deteriorates.

\subsection{Software Implementation}

The software tool developed by the author to implement the proposed method consists of the following parts: (1) a module for computing change metrics (listed in Table~\ref{tab:metrics}); (2) a module for finding edit scripts of code changes (based on the open-source version control system \textit{Subversion}); (3) a clustering module with partition quality functional optimization (based on the \textit{CLUTO} toolkit).

% ===========================================================================
\section{Conclusion}
\label{sec:conclusion}

The following results were obtained in this dissertation:

\begin{enumerate}
    \item The feasibility of partial automation of source code change classification by the metrics clustering method was justified.
    \item The choice of the \textit{k-means} method with a cosine-based object proximity measure for clustering, based on the cosine of the angle between change metric vectors, was justified.
    \item A method for automated classification of source code changes based on clustering of change metrics was developed, enabling a reduction in the number of changes that must be classified manually.
    \item Software implementing the proposed automated change classification method was developed.
\end{enumerate}

These results were obtained in the course of joint work at \textit{ITMO University} and \textit{Transas Technologies} and are used both in the development of complex software systems and in the educational process.

The work presents comparisons of change classifications for open-source software systems performed using the proposed automated method and manually. Independent experts with at least five years of experience in developing complex software systems were engaged for this purpose. In total, five software systems were analyzed. The research showed that the method enables a reduction in the time spent on change classification compared to ``purely manual'' expert review (for the NHibernate system, only 73 out of 2069 changes needed to be manually classified) with the following quality characteristics: purity $P_C = 0.75 \pm 0.05$ and entropy $E_C = 0.37 \pm 0.06$.

Currently, the method uses 11 metrics, listed in Section~\ref{sec:metrics}. The author suggests that further improvement of classification quality is possible by increasing the number of metrics used, as well as by adapting the method to other sets of change classes beyond those considered in the dissertation.

% ===========================================================================
\section*{Research Methods}

The work employed methods of cluster analysis, mathematical statistics, and software engineering.

\section*{Practical Significance}

All obtained results are currently in use and will continue to be used for improving software quality during the development of complex software systems. The proposed approach was applied for classifying source code changes in products developed by \textit{Transas Technologies} (mobile object monitoring system \textit{Navi-Manager}, a set of components of the global \textit{LRIT} system for tracking vessel positions worldwide, the student action monitoring system \textit{e-Tutor 5000}), as well as in two open-source software systems.

% ===========================================================================
\section*{Acknowledgments}

The original research was conducted at Saint Petersburg State University of Information Technologies, Mechanics and Optics (ITMO University) under the supervision of Prof.\ Anatoly A.\ Shalyto, D.Sc. The author gratefully acknowledges the contributions of Danil G.\ Shopyrin as co-author of the underlying publications~\cite{Knyazev2008a,Knyazev2008b,Knyazev2007a} and Prof.\ Shalyto's guidance throughout the doctoral research. The work was carried out in collaboration with Transas Technologies (Saint Petersburg).

% ===========================================================================

\end{document}